\documentstyle[psfig]{mn}

\input epsf.tex

\newcommand {\Mpc}   {\mbox{$h^{-1}$ Mpc \,}}

\newcommand{\mincir}{\raise -2.truept\hbox{\rlap{\hbox{$\sim$}}\raise5.truept
\hbox{$?$}\ }}
\newcommand{\gr}{\kern 2pt\hbox{}^\circ{\kern -2pt K}} 
\newcommand{\magcir}{\raise -2.truept\hbox{\rlap{\hbox{$\sim$}}\raise5.truept
\hbox{$?$}\ }}

\newcommand{\Om}{\Omega}
\newcommand{\be}{\begin{equation}}
\newcommand{\ee}{\end{equation}}
\newcommand{\bea}{\begin{eqnarray}}
\newcommand{\eea}{\end{eqnarray}}

\newcommand{\etal}{{et al.}}
\newcommand{\apj}{ApJ}

\newcommand{\mnras}{MNRAS}

\begin{document}

\title{Cosmological parameters from complementary observations of the Universe}

\author[R.~Durrer, B.~Novosyadlyj]
       {R.~Durrer$ ^1$,  B.~Novosyadlyj$ ^2$\\
$^1$ Department de Physique Th\'eorique, Universit\'e de Gen\`eve,
Quai Ernest Ansermet 24, CH-1211 Gen\`eve 4, Switzerland \\
$^2$ Astronomical Observatory of National University of L'viv, Kyryla and
Mephodia str.8, 290005, L'viv, Ukraine}

 

\maketitle

\begin{abstract}
We use observational data on the large scale structure (LSS) of the Universe
 measured over a wide range of scales from sub-galactic up to horizon scale 
and on the cosmic microwave background anisotropies
to determine cosmological parameters within the class of adiabatic
inflationary models. We show that a mixed dark matter model with
 cosmological constant ($\Lambda$MDM  model) and  parameters  
$\Omega_m=0.37^{+0.25}_{-0.15}$,    
$\Omega_{\Lambda}=0.69^{+0.15}_{-0.20}$,    
$\Omega_{\nu}=0.03^{+0.07}_{-0.03}$,    $N_{\nu}=1$,
$\Omega_b=0.037^{+0.033}_{-0.018}$, 
$n_s=1.02^{+0.09}_{-0.10}$,    
$h=0.71^{+0.22}_{-0.19}$,    
$b_{cl}=2.4^{+0.7}_{-0.7}$
 (1$\sigma$ confidence limits)    
matches  observational data on  LSS, the nucleosynthesis constraint, direct
measurements of Hubble constant, the  high redshift supernova type Ia 
results and the recent measurements of the location and amplitude of
the first acoustic peak in the CMB anisotropy power spectrum.
The best model is $\Lambda$ dominated (65\% of the total energy density)
and has slightly positive curvature, $\Omega=1.06$. The 
clustered  matter consists in 8\%  massive neutrinos, 10\%  baryons
and 82\%  cold dark matter (CDM). The upper 2$\sigma$ limit
on the neutrino content can be expressed in the form 
$\Omega_{\nu}h^2/N_{\nu}^{0.64}\le0.042$ or, via the neutrino mass, 
$m_{\nu}\le4.0$eV. The upper 1(2)$\sigma$ limit for the contribution
of a tensor mode to the COBE DMR data is T/S$<1(1.5)$.
Furthermore, it is shown  that the LSS observations 
together with  the Boomerang (+MAXIMA-1) data on the first acoustic
peak  rule out zero-$\Lambda$ models at more than $2\sigma$
confidence limit.

\end{abstract}

\begin{keywords}
Cosmology: large scale structure -- microwave background anisotropies
-- cosmological models: power spectrum -- cosmological parameters
\end{keywords}


\section{Introduction}

In the last decade of this century we have obtained  important
experimental results which play a crucial role for cosmology:
the COsmic Background Explorer has discovered the large scale anisotropies
of the cosmic microwave background radiation~\cite{ben96}; the
High-Z Supernova  Collaboration~\cite{rie98} and  the Supernova Cosmology 
Project~\cite{per98} found that the universe is accelerating rather
than decelerating; the
Super-Kamiokande experiment \cite{fuk98} discovered neutrino oscillations
which prove the existence of neutrinos with non-zero rest mass;
balloon-borne measurements of the cosmic microwave background (CMB) 
temperature fluctuations by Boomerang~\cite{ber00}
and MAXIMA-1~\cite{han00} have measured the height, position and width
of the first acoustic peak which is in superb agreement with an
adiabatic scenario of galaxy formation.

On the other hand the  comparison of recent experimental
data on the large scale structure of the Universe
with theoretical predictions of inflationary cosmology have shown
since quite some time that the simplest cold dark matter (CDM) model
is ruled out and we have to allow for a wider set of parameters to fit
all observational data on the status and history of our Universe. 
These include spatial curvature ($\Omega_k$), a 
cosmological constant ($\Omega_{\Lambda}$), the Hubble parameter
($h\equiv H_0/(100$km/s/Mpc$)$), the energy density of baryonic matter
($\Omega_b$),  cold dark matter ($\Omega_{cdm}$), the number of
species of massive neutrinos ($N_{\nu}$) and their density ($\Omega_{\nu}$),
the amplitude of the power spectra of primordial perturbations in
scalar ($A_s$) and tensor ($A_t$) modes and the corresponding
power-law indices  ($n_s$ and $n_t$), and the  optical depth to early
reionization ($\tau$). Constraining  this multidimensional parameter
space, determining the true values of fundamental cosmological
parameters, the nature and content of the matter which fills our
Universe is an important and exciting  problem of cosmology which has
now become feasible due to the enormous progress in cosmological observations. 
About a dozen or more papers have been devoted to this
problem in the last couple of years (see e.g.
\cite{lin97,lin98,efs99,Teg99,bri99,Mel00,nov00a,teg00a,teg00b,lyt00,nov00b,lan00,bal00,hu00,teg00c},  
some reviews are~\cite{dur99,sah99,pri00a,pri00b} and references
therein).

However, in spite of this intensive investigations the problem 
is still not satisfactorily resolved. Some of the remaining issues are
explained below.

First of all, we would like to have observations which
'measure' cosmological parameters in an as model independent way as
possible. Clearly, most values of cosmological parameters obtained
from observations of large scale structure, galaxy clustering and CMB
anisotropies are strongly model dependent. If the 'correct'
model of structure formation is not within the family investigated, we may not
notice it, especially if the error bars are relatively large. This
leads us to the next problem. Even if cosmological observation have
improved drastically, we still need more accurate data with better
defined statistical properties (e.g we need to know the correlation of
different measurements). The new CMB anisotropy data are already of
this quality but the galaxy and cluster data are still relatively far
from it.

 A next important point is the correspondence between
theoretical predictions and observational characteristics
used in the analysis. We have to find a fast but accurate way to
compute the theoretical values, especially when exploring high
dimensional parameter spaces.
All parameters must be fitted simultaneously which renders the 
problem computationally complicated and very time consuming. Due to
this difficulty, many authors search some subset of parameters setting
the others to some fixed 'reasonable' priors, thereby investigating
a sub-class of cosmological models. As different authors also use
different subsets  of observational data, the resulting cosmological
parameters still vary in a relatively wide range. 

Another problem  are the degeneracies in parameter space which appear
especially  in the case when only CMB anisotropy data are
used~\cite{efs99}. It can be reduced substantially or even removed
completely if galaxy clustering data, corresponding to different scales
and redshifts, are combined with CMB measurements. This idea has
already been employed on several occasions and is known under the name
'cosmic concordance' (for a recent review see Tegmark \etal~(2000c)).

The goal of this paper is to determine  cosmological parameters of the
sub-class of models without tensor mode and no early reionization on
the basis of LSS data related to different scales and different
redshifts. In Novosyadlyj \etal~(2000a) we have used  the same approach 
to test flat models; we have shown that $\Lambda$MDM models
are preferred in this class of models. There we have also shown  that
pure CDM models with $h\ge 0.5$, scale invariant 
primordial power spectrum, vanishing cosmological constant and spatial
curvature are  ruled out at very high confidence level, more than
$99.99\%$.  The corresponding class of mixed dark matter (MDM) models
are ruled out at  about $95\%$ C.L. It was noted~\cite{nov00b} that
the galaxy clustering data set determines the amplitude of scalar
fluctuations approximately at the same level as the COBE four-year
data. This indicates that a possible tensor component in the COBE data
cannot be  very substantial. 

In this paper we test $\Lambda$MDM models with non-zero
curvature.  Furthermore,  we use the data on the
location and amplitude of the first acoustic peak determined from
the most accurate recent measurements of the CMB power spectrum.
The data on the amplitude of the 2-nd and 3-d peaks is used as additional test
for model prefered by large scale structure,  
COBE and 1-st peak data. We investigate the {\mbox (in-)consistency} of our
data set with the 2-nd and 3-d peaks.
We also use the  SNIa constraint for comparison.

The outline of the paper is as follows: in Sect.~2 
we describe the experimental data set which is used here.
The calculations of theoretical  
predictions and the method employed to determine cosmological parameters
are described in Sect.~3. 
In Sect.~4 we discuss our results and compare them with other investigations.
Our conclusions are presented in Sect.~5. 

\section{The experimental data set}

Our approach is based on the quantitative comparison of the theoretical
predictions for the characteristics of the large scale structure of the Universe 
with corresponding observational ones. Theoretical predictions are calculated
on the base of initial power spectrum of density perturbations of which shape
strongly depends on all parameters supposed here for determination. 
So, model independent observational constraints on the inclination and 
amplitude of the power spectrum at different scales will be used in this search.

\subsection{CMB data}

We use the COBE 4-year data on CMB
temperature anisotropies \cite{ben96} to normalize the density fluctuation 
power spectra according to Liddle \etal~(1996) and Bunn \&~White~(1997).
Therefore, each model will match the COBE data by construction. 

We believe that using all
available experimental data on $\Delta T/T$ at angular scales smaller
than the COBE measurement is not an optimal way to search
best-fit cosmological parameters due to their large dispersion 
(see for examples Fig.10.1 of Durrer \& Straumann 1999, Fig.2 of
Novosyadlyj \etal~ 2000a or Fig.1 of Tegmark \etal~2000)
which together with the large number of experimental points, $\sim70$  
stipulating a high degrees of freedom, result in wide ranges for the
confidence limits on cosmological parameters. 
The Boomerang \cite{ber00} and MAXIMA-1 \cite{han00} experiments represent 
a new generation of CMB measurements. They have produced a CMB map of about 
$\sim$100deg$^2$  with a resolution better than half a degree and a
S/N$\sim$2, which  allows to determine the location and amplitude of
the first acoustic peak with high accuracy.
The position of the first and amplitudes of the
first, second and third acoustic  peaks in the angular power spectrum of the 
CMB temperature fluctuations together with the COBE data 
are the main measured characteristics of the CMB power
spectrum. They contain information about amplitude and tilt of the
primordial power  spectrum of
density fluctuations at largest scales, from  few tens of Mpc up to
the current horizon scale of several thousand Mpc. They are mainly
sensitive to the parameters 
$\Omega_k$, $\Omega_mh^2$, $\Omega_{\Lambda}$, $\Omega_bh^2$,
$n_s$ and to the normalization of the initial power spectrum of density
fluctuations. 

 For example, the
Boomerang data indicate that the first peak is located at the Legendre
multipole $\tilde\ell_p=197\pm6$ and has an amplitude of $\tilde
A_p=69\pm8\mu \rm K$ (this
1$\sigma$ error includes statistical and calibration errors). 
Here and in the following a tilde denotes observed quantities.
 We use these results in our search procedure.
The MAXIMA-1 data ($\tilde\ell_p\approx 220$, $\tilde A_p=78\pm6\mu \rm K$) 
marginally match Boomerang data and we will show that using them in combination
with Boomerang data does not change the results significantly. The positions
of 2-nd and 3-d peaks are not well determined and we will not use
them in the main search procedure but we use their amplitudes as
determined by \cite{hu00}
for comparison with the predictions of our best-fit model.

\subsection{Rich cluster data}

The important constraints on the form and amplitude of the matter
power spectrum in the range
from $10h^{-1}$Mpc up to scales approaching $1000h^{-1}$Mpc can be obtained 
from the study of clusters of galaxies, their space distribution,
mass and X-ray temperature functions. 

The power spectrum reconstructed from the observed space distribution of 
clusters has been determined many times for different samples from Abell, 
ACO and APM catalogs (see Einasto et al. 1997, Retzlaff et al. 1998,
Tadros et al. 1998, Miller and Batuski 2000 and references therein). 
The remarkable feature of the determinations by different groups is 
similar slopes of cluster power spectra on scales $0.02h{\rm Mpc^{-1}}
\le k\le 0.1h{\rm Mpc^{-1}}$,
$n\sim -1.5$ (see above mentioned references). 
Here, we use the power spectrum of Abell-ACO clusters
 $\tilde P_{A+ACO}(k_j)$ \cite{ret97} as observational
input. It is measured in the range $0.03h/$Mpc$\le k\le 0.2h/$Mpc
where effects of nonlinear evolution are negligible and it has well analyzed
sources of uncertainties. The cluster power spectrum is biased with
respect to the dark matter distribution.  We assume that the bias is
linear and scale independent. This is reasonable in the range of
scales considered as predicted from local bias models~\cite{fg93} and
indicated by numerical simulations~\cite{bcfbl00}. In our previous paper 
\cite{nov00a} we have shown that not all the 13 points given in
Retzlaff et al. (1998) are independent
measurements and the effective number of degrees of freedom is 3. But 
to make best use of the observational information we use all 13 points
of the power spectrum to determine cosmological parameters and assign
$n_F=3$ for the number of degrees  of freedom in the marginalization procedure.

\medskip

A constraint for the amplitude of the fluctuation power spectrum on
cluster scale can be derived from the cluster mass and the X-ray temperature
functions. It is usually formulated as constraint for the density
fluctuation in a top-hat sphere of 8\Mpc radius, $\sigma_{8}$, which
can be calculated for a given initial power spectrum $P(k)$ by
\be
\sigma_{8}^{2}={1\over
2\pi^{2}}\int_{0}^{\infty}k^{2}P(k)W^{2}(8{\rm Mpc}\;k/h)dk,
\label{si8}
\ee
where $W(x)=3(\sin x-x \cos x)/x^3$ is the Fourier transform of a
top-hat window function.
Recent optical determinations of the mass function of nearby galaxy
clusters \cite{gir98} give
\be
\tilde\sigma_{8}\Omega_m^{\alpha_1}=0.60\pm 0.04
\ee
where $\alpha_1=0.46-0.09\Omega_m$ for flat low-density models and
$\alpha_1=0.48-0.17\Omega_m$ for open models (at the 90\% C.L.). 
Several groups have found similar results using  different
methods and different data sets (for a comprehensive list of references
see Borgani \etal~(1999)). This constraint on $\sigma_8$ is
exponentially sensitive and thus allows only very small error bars. If
the theory is correct this is of course a great advantage. However, if
our understanding of cluster formation is not entirely correct, this
will lead to discrepancies with other experimental constraints.

From the observed evolution of the cluster X-ray temperature
distribution function between $z=0.05$ and $z=0.32$ we use the
following constraint derived by Viana \&~Liddle (1999)
$$\tilde\sigma_8\Omega_m^{\alpha_2}=0.56\pm0.19\Omega_m^{0.1\lg\Omega_m+\alpha_2},
\;\;\;\alpha_2=0.34$$
for open models and
$$\tilde\sigma_8\Omega_m^{\alpha_2}=0.56\pm0.19\Omega_m^{0.2\lg\Omega_m+\alpha_2},
\;\;\;\alpha_2=0.47$$
for flat models (both with 95\% confidence limits).

From the existence of three very massive clusters of galaxies observed
so far at $z>0.5$ an additional constraint has been established by
\cite{bah98}
\be
\tilde \sigma_8\Omega_m^{\alpha_3}=0.8\pm 0.1\;,
\ee
where $\alpha_3=0.24$ for open models and $\alpha_3=0.29$ for flat
models.
 
Note that all these constraints are given by slightly different
formulas for either $\Omega_\Lambda=0$ or $\Omega_\Lambda +
\Omega_m=1$. However, we are going to use them for arbitrary values of
$\Omega_\Lambda$ and $\Omega_m$. Since our best fit models are
relatively close to the flat model, we mainly use the formula for the
flat case. We have checked that our results are insensitive to this choice.

\subsection{Peculiar velocity data} 

Since our approach is based on the initial power spectrum of density 
fluctuations it seems most favorable to use the power spectrum reconstructed 
from the observed space distribution of galaxies. But the galaxy power spectra 
obtained from the two-dimensional  APM survey 
(e.g. \cite{mad96,tad96}, and references therein), the CfA redshift survey 
\cite{vog92,par94}, the IRAS survey \cite{sau92,sau00} and the Las 
Campanas Redshift Survey \cite{dc94,lan96} differ significantly in both, the 
amplitude and the position of the maximum.  Moreover, nonlinear 
effects on small scales must be taken into account in their analysis. 
On the other hand, these power spectra contain large number of experimental
points which are not independent and a decorrelation
procedure for these power spectra must be employed. 
For these reasons and also in order to test the consistency between
different data set, we do not include galaxy power spectra for the 
determination of parameters in this work. It will be interesting to compare
our best fit parameters with those obtained in analyzes
including galaxy power spectra.  

Another constraint on the amplitude of the linear power spectrum of 
density fluctuations in our vicinity comes from the study of  
bulk flows of galaxies in spheres of large enough radii around our 
position. Since these data may be influenced by the local super-cluster 
(cosmic variance), we will use only the value of the bulk motion - the 
mean peculiar velocity of galaxies in a sphere of radius 
$50h^{-1}$Mpc given by \cite{kol97}, 
\be 
\tilde V_{50}=(375\pm 85) {\rm km/s.} 
\ee 
With its generous error bars, this value is in a good agreement with
other measurements of bulk motion at 
the scale $40-60h^{-1}$Mpc \cite{ber90,cou93,dek94} (see also the
review by Dekel 1999).

\subsection{Ly-$\alpha$ constraints} 

An important constraint on the linear matter power spectrum
on small scales ($k\sim (2-40)h/$Mpc) comes
from the Ly-$\alpha$ forest, the Ly-$\alpha$ absorption lines seen in
quasar spectra (see Gnedin (1998), Croft \etal~(1998)
 and references therein).  Assuming that the
Ly-$\alpha$ forest is formed by discrete clouds with a physical size close
to the Jeans scale in the reionized inter-galactic medium at $z\sim 2-4$,
Gnedin (1998) has derived a constraint on the value of the
r.m.s. linear density fluctuations
\bea
 1.6<\tilde \sigma_{F}(z=3)<2.6~~(95\% \mbox{C.L.}) &&\\
 \mbox{ at }~ k_{F}\approx 34\Omega_m^{1/2}h/{\rm Mpc}~. \nonumber
\eea
Taking into account the
new data on quasar absorption lines, the effective equation of state
and the temperature of the inter-galactic medium at high redshift
were re-estimated
recently~\cite{ric99}. As a result the value of Jeans scale at $z=3$ has
moved to $k_{F}\approx 38\Omega_m^{1/2}h/$Mpc \cite{gn99}. Here, we adopt this 
new value.

The procedure to
recover the linear power spectrum from the Ly-$\alpha$ forest has been
elaborated by Croft \etal (1998). Analyzing the absorption lines in a sample
of 19 QSO spectra, they have obtained the following constraint on the
amplitude and slope of the linear power spectrum at $z=2.5$ and
$k_{p}=1.5\Om_m^{1/2}h/$Mpc,
\be
\tilde \Delta_{\rho}^2(k_p)\equiv k_p^3P(k_p)/2\pi^2=0.57\pm 0.26,
\ee
\be
\tilde n_p\equiv {\Delta \log\;P(k)\over \Delta \log\;k}\mid
_{k_p}=-2.25\pm 0.18,
\ee
at (1$\sigma$ C.L.). 
The like constraints on the amplitude and slope of the linear power spectrum
was obtained by \cite{mcd00} from the analysis of absorption lines in a sample 
of 8 QSO. We will analyze these constraints in the context of our task and 
compare them with previous two. But in the main search procedure we will use the  
constraints given by Croft \etal (1998) as based on the more extensive sample
of quasars.

\subsection{Other experimental constraints} 

In addition to the CMB \& LSS measurements described above we also use
some results of global observations which are independent of the LSS model.
For the value of the Hubble constant we set
\be
\tilde h=0.65\pm 0.10, \label{Hubble}
\ee
which is a compromise between measurements made by two groups,
\cite{tam97} and \cite{mad98}. We also employ a nucleosynthesis
constraint on the baryon density deduced from the determination 
of the primeval deuterium abundance 
\be
\widetilde{\Omega_bh^2} = 0.019\pm 0.0024 ~~~ (95\% {\rm C.L.})
\ee
given by Burles \etal~(1999). The new, more precise determination
\cite{bur00} confirms this value.

Furthermore, we include the distance measurements of super novae
of type Ia (SNIa) which constrain the cosmic expansion
history~\cite{rie98,per98,per99}). In a universe with cosmological
constant this gives an important 
 constraint on a combination of the values of the curvature, the
cosmological constant and the matter content of the Universe. We use
the following constraint in our parameter search~\cite{per99}
\be
\widetilde{[\Omega_m-0.75\Omega_{\Lambda}]}=-0.25\pm0.125~.
\ee

\section{The method and some tests }

One of the main ingredients for the solution for our search problem
is a reasonably fast and accurate determination of the  linear
transfer function for dark matter clustering
which depends on the cosmological parameters. We use accurate analytical
approximations of the MDM transfer function $T(k;z)$ depending on
the parameters $\Omega_m$, $\Omega_b$, $\Omega_{\nu}$, $N_{\nu}$ and
$h$ by Eisenstein \&~Hu (1999). According
to this work, the
linear power spectrum of matter density fluctuations is given by
\be
P(k;z)=A_sk^{n_s}T^2(k;z)D_1^2(z)/D_1^2(0),\label{pkz}
\ee
where $A_s$ is the normalization constant for scalar perturbations and
$D_1(z)$ is the linear growth factor, which can be approximated by
\cite{car92} 
$$D_1(z)=
{5\over 2}{\Omega_m(z)\over 1+z}\left[{1\over
70}+{209\Omega_m(z)-\Omega_m^2(z) \over 140}+\Omega_m^{4/7}(z)\right]^{-1},$$
where
$\Omega_m(z)=\Omega_m(1+z)^3/\left(\Omega_m(1+z)^3+\Omega_{\Lambda}
		+\Omega_k(1+z)^2 \right)$.

We normalize the spectra to the 4-year COBE data
which determine the amplitude of density perturbation at horizon
scale, $\delta_h$ \cite{lid96,bun97}.
 The normalization constant $A_s$ is then given by
\be
A_s=2\pi^{2}\delta_{h}^{2}(3000{\rm Mpc}/h)^{3+n_s}
  ~ .\label{anorm}
\ee

The Abell-ACO power spectrum is related to the matter power
spectrum at $z=0$, $P(k;0)$, by the cluster biasing parameter $b_{cl}$.
As argued above, we assume scale-independent, linear bias
\be
P_{A+ACO}(k)=b_{cl}^{2}P(k;0).\label{pcl}
\ee
For a given set of parameters  $\Omega_m$, $\Omega_{\Lambda}$, $\Omega_b$,
$\Omega_{\nu}$, $N_{\nu}$, $n_s$, $h$,
 and $b_{cl}$  the theoretical values of
$P_{A+ACO}(k_j)$ can now be obtained for the values $k_j$ 
(Table 1 of \cite{nov00a}). We denote them by $y_j$ ($j=1,...,13$).

The dependence of the position and amplitude of the first acoustic
peak in the CMB power spectrum on cosmological
parameters has been investigated using CMBfast~\cite{sz96}.  
As expected, and  as we have shown in our previous
paper~\cite{nov00a}, the results are, within reasonable
accuracy, independent on the fraction of hot dark matter,
$f_{\nu}=\Om_\nu/\Om_m$, up to $f_{\nu}~0.6$.

For the remaining parameters, $n_s$, $h$, $\Omega_b$, $\Omega_{cdm}$ and
$\Omega_{\Lambda}$, we determine the resulting values $\ell_p$ and
$A_p$ using the analytical approximation given by Efstathiou \&~Bond (1999).
We extend the approximation to  models with non-zero curvature 
($\Omega_k\equiv1-\Omega_m-\Omega_{\Lambda}\ne0$) by adding a
coefficient for the amplitude and  the peak location, which is determined 
numerically. 
The analytical approximation for the position of 
the first acoustic peak used here is
\be
\ell_p=0.746\pi\sqrt{3(1+z_r)}{R(\omega_m,\omega_k,y)\over 
I_s(\omega_m,\omega_b)},
\ee
where $\omega_*\equiv\Omega_*h^2$, and 
$R=\omega_m^{1/2}{\sinh(\omega_k^{1/2}y)\over 
\omega_k^{1/2}}$, $\omega_m^{1/2}y$, $\omega_m^{1/2}
{\sin(\mid \omega_k\mid^{1/2}y)\over \mid\omega_k\mid^{1/2}}$
for open, flat and closed models respectively. Here 
$y(\omega_m,\omega_k,\omega_{\Lambda})$ is given by formula (8b) and 
$I_s(\omega_m,\omega_b)$ by formulae (17-19) of~\ Efstathiou \&~Bond (1999).
The accuracy of this analytical approximation is better than 1\%.

The approximation for the amplitude of first acoustic peak 
is as follows:
\bea
A_p &=&\left({\ell_p(\ell_p+1)\over 2\pi}C_2{\Gamma(l_p+{n_s+1\over2})
       \over\Gamma(l_p+{5-n_s\over2})}
       {\Gamma({9-n_s\over2})\over\Gamma({3+n_s\over2})} \right. \nonumber \\
 &&     + 0.838A(\omega_b,\omega_{cdm},n_s)\big)^{1/2},
\eea
where $\ln A(\omega_b,\omega_{cdm},n_s)=4.5(n_s-1)+a_1+a_2\omega_{cdm}^2+
a_3\omega_{cdm}+a_4\omega_b^2+a_5\omega_b+a_6\omega_b\omega_{cdm}+a_7\omega_k$,
with $a_1=2.376$, $a_2=3.681$, $a_3=-5.408$, $a_4=-54.262$, $a_5=18.909$,
$a_6=15.384$, $a_7=4.2$ and $C_2$ is the quadrupole anisotropy
approximated by
\be
C_2=A_s{\pi\over 16}\left({H_0\over c}\right)^{n_s+3}
{\Gamma(3-n_s)\over\Gamma^2({4-n_s\over2})}
{\Gamma(2+{n_s+1\over2})\over\Gamma(2+{5-n_s\over2})}.
\ee
The values $a_1-a_6$ are the best-fit coefficients determined from a
grid of models computed with CMBfast~\cite{efs99}. We have added the
coefficient  $a_7$ in order to account for curvature. The
accuracy of $A_p$ in the parameter ranges
which we consider is better than 5\%. We
denote $\ell_p$ and $A_p$ by $y_{14}$ and $y_{15}$ respectively.

The theoretical values of the other experimental constraints are
obtained as follows: the density fluctuation $\sigma_8$ is
calculated according to Eq.~(\ref{si8}) with $P(k;z)$ taken from
Eq.~(\ref{pkz}). We
set $y_{16}=\sigma_{8}\Omega_m^{\alpha_1}$, 
$y_{17}=\sigma_{8}\Omega_m^{\alpha_2}$ and $y_{18} =
\sigma_{8}\Omega^{\alpha_3}_m$
with corresponding values of $\alpha_i$ (i=1, 2, 3) for vanishing and
non-zero curvature (see previous section).

The r.m.s. peculiar velocity of
galaxies in a sphere of radius $R=50h^{-1}$Mpc is given by
\be
V^{2}_{50}={1\over 2\pi^{2}}
\int_{0}^{\infty}
k^2P^{(v)}(k)e^{-k^{2}R_{f}^{2}}W^{2}(50{\rm Mpc}~k/h)dk,  \label{V50th}
\ee
where $P^{(v)}(k)$ is power spectrum for the velocity field of the
density-weighted matter~\cite{eh3},
$W(50{\rm Mpc}~k/h)$ is the top-hat window function.
A previous smoothing
of the raw data with a Gaussian filter of radius $R_{f}=12h^{-1}$Mpc is
employed, similar to the procedure which has led to the observational
value.  For the scales of interest
 $P^{(v)}(k)\approx (\Omega^{0.6}H_0)^2P(k;0)/k^2$.
We denote the r.m.s. peculiar velocity by $y_{19}$.

The  value by Gnedin (1998) from the formation of Ly-$\alpha$ clouds
constrains the r.m.s. linear density perturbation at redshift $z=3$ and
wave number $k_{F}=38\Omega_m^{1/2}h/$Mpc. In terms of the power
spectrum,  $\sigma_F$ is given by
\be
\sigma_{F}^{2}(z)={1\over
2\pi^{2}}\int_{0}^{\infty}k^{2}P(k;z)e^{(-k/k_F)^2}dk,\label{siF}.
\ee
It will be denoted by $y_{20}$. The corresponding value of the
constraint by Croft \etal~(1998) is
\be
\Delta_{\rho}^2(k_p,z)\equiv k_p^3P(k_p,z)/2\pi^2,   \label{Dekp}
\ee
\vbox{at $z=2.5$ with $k_{p}=0.008H(z)/(1+z)({\rm km/s})^{-1}$,
 ~will~ be } \vbox{denoted by $y_{21}$;
$H(z)=H_0\left[\Omega_m(1+z)^3+\Omega_k(1+z)^2+\Omega_{\Lambda}\right]^{1/2}$ }
is the Hubble parameter at redshift z. The slope
of the power spectrum at this scale and redshift,
\be
n_p(z)\equiv {\Delta \log\;P(k,z)\over \Delta \log\;k}~,\label{enp}
\ee
is denoted by $y_{22}$.

For all tests except Gnedin's Ly-$\alpha$ clouds,
we use the density weighted  transfer function $T_{cb\nu}(k,z)$
from \cite{eh3}. For Gnedin's $\sigma_F$ we use $T_{cb}(k,z)$
according to the prescription of Gnedin~(1998). It must be noted that
even in the model with maximal $\Omega_{\nu}$ ($\sim0.2$) the difference
between $T_{cb}(k,z)$ and $T_{cb\nu}(k,z)$ is less than $12\%$ for  $k\le k_p$.
Early reionization changes  somewhat the evolution of density
perturbation in the baryon component on small scales. This effect is
not taken into account by the analytical approximation used here
\cite{eh3}. Therefore, we restrict ourselves to models without 
early reionization. We calculate the Ly-$\alpha$ tests according to
the prescription given in Sect. 5.4 of \cite{eh3}.

Finally, the values of $\Omega_bh^2$, $h$ and
$\Omega_m-0.75\Omega_{\Lambda}$ are denoted by $y_{23}$,
$y_{24}$ and $y_{25}$ respectively.

The squared differences between the
theoretical and observational values divided by the observational
error are given by $\chi^2$,
\be
\chi^{2}=\sum_{j=1}^{23}\left({\tilde y_j-y_j \over \Delta \tilde y_j}
\right)^2~.     \label{chi2}
\ee
Here $\tilde y_j$ and $\Delta \tilde y_j$ are the experimental data
and their dispersion, respectively. The set of parameters 
$\Omega_m$, $\Omega_{\Lambda}$, $\Omega_{\nu}$, $N_{\nu}$, $\Omega_b$, 
$h$, $n_s$ and $b_{cl}$ are then determined by minimizing $\chi^2$ using the
Levenberg-Marquardt method~\cite{nr92}. The derivatives of the predicted
values with respect to the search parameters which are required by this method
are obtained numerically using a relative step size of $10^{-5}$ with respect
to the given parameter.

In order to test our method  for stability, we have constructed a mock
sample of 
observational data. We start with a set of cosmological parameters and
determine the ``observational'' data for them which would be measured
in case of faultless measurements with $1\sigma$ errors comparable to
the observational errors.  We then insert random sets of starting
parameters into the search program  and try to recover the model
which corresponds to the mock data. The method is stable if we can
recover our input cosmological model (for more details of this test procedure
see~Novosyadlyj \etal~(2000a). The code finds all the previously known
parameters with high accuracy. Even starting very far away from
the true values, our method reveals as very stable and finds the
'true' model whenever possible. This means that the code finds the
global minimum of $\chi^2$ independent of the initial values for the
parameters. This also hints that our data set is sufficiently divers
to be free of degeneracies (which plague parameter searches working
with CMB data only).

\section{Results and Discussion}

\subsection{Calculations}

The determination of the parameters 
\footnote{We treat $\Omega_{\Lambda}$ and $\Omega_m$  as free
parameters, $\Omega_k=1- \Omega_{\Lambda}- \Omega_m$.}
$\Omega_m$, $\Omega_{\Lambda}$, $\Omega_{\nu}$, $N_{\nu}$, $\Omega_b$, 
$h$, $n_s$ and $b_{cl}$ by the Levenberg-Marquardt $\chi^2$
minimization method~\cite{nr92} can be realized in the following way: we vary the
set of parameters 
$\Omega_m$, $\Omega_{\Lambda}$, $\Omega_{\nu}$, $\Omega_b$, 
$h$, $n_s$ and $b_{cl}$ with fixed $N_{\nu}$ and find the
minimum of
$\chi^2$. Since $N_{\nu}$ can be discrete we repeat this
procedure three times for $N_{\nu}$=1, 2, and 3.  The lowest of the
three minima is the minimum of $\chi^2$ for the
complete set of free parameters. 
Hence, we have seven free parameters.
The formal number of observational points is 25 but, as we have mentioned,  the 13
power spectra points can be described by just 3 degrees of freedom, so that the
maximal number of truly independent measurements is 15. Therefore, 
the number of degrees of
freedom for our search procedure is $N_F= N_{\rm exp}-N_{\rm par}= 8$ 
if all observational points are used. In order to investigate to what extent the
LSS constraints on fundamental parameters match the constraints implied by
SNIa \cite{per99} we have determined all 8 parameters with and 
without the SNIa constraint ($\tilde y_{25}$). The results are presented in the 
Table~\ref{rez1}. 

Note, that
for all models $\chi^2_{min}$ is in the range
$N_F-\sqrt{2N_F}\le \chi^2_{\min}\le N_F+\sqrt{2N_F}$ which is
expected for a Gauss\-ian distribution of $N_F$
degrees of freedom. This means that the cosmological paradigm which
has been  assumed is in agreement with the data.
(Note here, that the reduction of the 13 not
independent data points of the cluster power spectrum to three parameters is not
important for our analysis since removing them from search procedure does not
change the results essentially, as we will see later.)

Let us investigate how the parameters of the best fit model vary if we
include also  the data of the MAXIMA-1 experiment. The location and
amplitude of the  first acoustic peak determined from the combined
Boomerang and MAXIMA-1 data 
are~\cite{hu00}  $\ell_p=206\pm6$, $A_p=78.6\pm7$. If we use them instead
values used above, the best fit parameters remain practically unchanged, 
$\Omega_m=0.37\pm0.06$, $\Omega_{\Lambda}=0.66\pm0.06$,
$\Omega_{\nu}=0.03\pm0.03$, $N_{\nu}=1$, $\Omega_b=0.039\pm0.010$,
$n_s=1.05\pm0.04$, and $h=0.70\pm0.09$. Hence, including the MAXIMA-1 data
into the determination of the first acoustic peak is not
essential in our analysis and we will use here the values determined from the
Boomerang data alone. This is however an important confirmation of
the consistency of the two data sets. 

We have  also analyzed the influence of the amplitudes of the 2-nd and
3-d acoustic peaks 
on the determination of cosmological parameters in the frame of our approach.
If we add to our data set their values and errors as determined
by~\cite{hu00} and calculate them using the analytical approximation
given by the same authors then $\chi^2\approx18$, which is far too
much for 9 degrees of freedom.  
In this case the best-fit parameters are
$\Omega_m=0.37\pm0.07$, $\Omega_{\Lambda}=0.72\pm0.05$, 
$\Omega_{\nu}\approx0$, $\Omega_b=0.046\pm0.011$, 
$n_s=0.97\pm0.03$, and $h=0.67\pm0.08$.
For the  2-nd acoustic peak and nucleosynthesis constraint the
deviations of the predicted values
from their observed counterparts are  maximal ($2.8\sigma$ higher, and
$1.4\sigma$ higher respectively). If we exclude the nucleosynthesis
constraint from the search procedure then $\chi^2/N_F\approx7/8$ 
and best-fit parameters become
$\Omega_m=0.34\pm0.06$, $\Omega_{\Lambda}=0.74\pm0.05$, 
$\Omega_{\nu}\approx0$, $\Omega_b=0.055\pm0.012$, 
$n_s=0.98\pm0.03$, and $h=0.72\pm0.08$. Practically all used constraints
are satisfied but $\Omega_bh^2$ is $9\sigma$ higher then value deduced
from the determination of the primeval deuterium abundance by \cite{bur99}
and $12\sigma$ higher then  more recent value~\cite{bur00}. 
This problem of the inconsistency of the Boomerang and MAXIMA-1 values
for the height of the second peak especially with the nucleosynthesis
constraint on the baryon abundance has been discussed at large in the
recent literature~\cite{lan00,teg00b,hu00,emmmp00,dur00}. Since we
have nothing new to add to this
subject here, we will not discuss it any further in this work. In what
follows, we exclude the 2-nd and 3-d acoustic peaks
from experimental data set in our search procedure but we will use them 
in the discussion of our best-fit model.

The errors in the best-fit parameters as presented in Table~\ref{rez1}
are the square roots of the  diagonal elements of the covariance
matrix which is calculated according to the prescription given in 
Press et al. 1992
(Chaper 15) or Tegmark \& Zaldarriaga 2000a (Appendix A). 

\subsection{The best-fit model}

The
model with one sort of massive neutrinos provides the best fit to
the data, $\chi^2_{min}=5.9$. 
However,  there is
only a marginal difference in $\chi^2_{min}$ for $N_\nu =1,2,3$.  
With the given accuracy of the data we cannot conclude
whether  massive neutrinos are present
at all and if yes what number of degrees of freedom is favored. 
We summarize, that the
considered observational data on LSS of the Universe can be
explained by a $\Lambda$MDM inflationary model with a scale invariant
spectrum of scalar perturbations and a small positive curvature.

Including of the SNIa constraint into the experimental data set
 decreases $\Omega_m$, increases $\Omega_{\Lambda}$ slightly and
prefers $\Omega_{\nu}\approx0$, a $\Lambda$CDM model.

In Table~\ref{constr} 
we compare the values of the different observational
constraints with the predictions for the best-fit models 
(Table~\ref{rez1}  for $N_{\nu}=1$). 
In both cases the calculated characteristics of the LSS
are within the $1\sigma$ error bars of the observed values. In the last row
we indicate the age of the Universe determined according to the general
expression for non-zero curvature and non-zero $\Lambda$  models~\cite{sah99}  
\be
t_0=H_0^{-1}\int^{\infty}_0{dz\over \Omega_m(1+z)^5+\Omega_k(1+z)^4+\Omega_{\Lambda}(1+z)^2}~.
\label{t0}
\ee
The predicted age of the Universe agrees well with recent 
determinations of the age of globular clusters.

\begin{table*}
\caption{Cosmological parameters determined from the LSS data
 described in the text without and with the SNIa constraint. The
 errors indicated are the square 
roots of the diagonal elements of the covariance matrix.\label{rez1}} 
\begin{center}
\def\onerule{\noalign{\medskip\hrule\medskip}}
\medskip
\begin{tabular}{|cccccccc|}
\hline
&&&&&&&\\
$N_{\nu}$     & $\chi^2_{min}$& $\Omega_m$& $\Omega_{\Lambda}$&$\Omega_{\nu}$& $\Omega_b$  &$n_s$   & $h$  \\ [4pt]
\hline
&&&&&&&\\
&&\multicolumn{4}{c}{Without SNIa constraint}&&\\
&&&&&&&\\
1&5.90&0.37$\pm$0.06&0.69$\pm$0.07&0.03$\pm$0.03&0.037$\pm$0.009&1.02$\pm$0.04&0.71$\pm$0.09\\
2&6.02&0.42$\pm$0.08&0.64$\pm$0.09&0.04$\pm$0.04&0.038$\pm$0.010&1.03$\pm$0.04&0.71$\pm$0.09\\
3&6.17&0.47$\pm$0.10&0.59$\pm$0.08&0.06$\pm$0.01&0.038$\pm$0.010&1.04$\pm$0.03&0.70$\pm$0.09\\
&&&&&&&\\
&&\multicolumn{4}{c}{Including SNIa constraint}&&\\
&&&&&&&\\
0-3&6.02&0.32$\pm$0.05&0.75$\pm$0.06&$<10^{-4}$&0.038$\pm$0.010&1.0$\pm$0.05&0.70$\pm$0.09\\
\hline
\end{tabular}
\end{center}
\end{table*}
\begin{table*}
\caption{Theoretical  predictions for the used characteristics  of the
best-fit  $\Lambda$MDM model with one sort of massive neutrinos with 
the cosmological parameters given in Table~\ref{rez1}, first line
(without SNIa constraint)  and last line (including the SNIa
constraint) are compared with observations. \label{constr} } 
\begin{center}
\def\onerule{\noalign{\medskip\hrule\medskip}}
\medskip
\begin{tabular}{|cccc|}
\hline
&&&\\
&&\multicolumn{2}{c}{Predictions}\\
\cline{3-4}
Characteristics& Observations$^{a)}$&Without SNIa &Including SNIa  \\ [4pt]
\hline
&&&\\
$\ell_p$                       &197$\pm$6        &197   &197   \\
$A_p$                          &69$\pm$8         &71.5  &71.9  \\
$V_{50}$, km/s                 &375$\pm$85       &327   &308   \\
$\sigma_8\Omega_m^{\alpha_1}$  &0.60$\pm$0.022   &0.61  &0.60  \\
$\sigma_8\Omega_m^{\alpha_2}$  &0.56$\pm$0.095   &0.58  &0.58  \\
$\sigma_8\Omega_m^{\alpha_3}$  &0.8$\pm$0.1      &0.69  &0.71  \\
$\sigma_F$                     &$2.0\pm .3$      &1.9   &1.9   \\
$\Delta^2_{\rho}(k_p)$         &$0.57\pm 0.26$   &0.56  &0.59  \\
$n_{p}(k_p)$                    &$-2.25\pm0.2$    &-2.20 &-2.20 \\
$h$                            &0.65$\pm$0.10    &0.71  &0.70  \\
$\Omega_bh^2$                  &0.019$\pm$0.0012 &0.019 &0.019 \\
$\Omega_m-0.75\Omega_{\Lambda}$&-0.25$\pm$0.125  &-0.14 &-0.25 \\
$t_0$, Gyrs$^{b)}$ &$13.2\pm 3.0$$^{c)}$, $11.5\pm 1.5$$^{d)}$     &12.6  &13.5  \\
\hline
\end{tabular}
\end{center}
$ ^{a)}$ all errors are $\pm 1\sigma$, 
$ ^{b)}$ is not used in the search procedure, 
$ ^{c)}$\cite{car99}, 
$ ^{d)}$\cite{cha98}
\end{table*}
\begin{table*}
\caption{Best-fit values of cosmological parameters determined from the
different data set.\label{dds} } 
\begin{center}
\def\onerule{\noalign{\medskip\hrule\medskip}}
\medskip
\begin{tabular}{|ccccccccc|}
\hline
&&&&&&&&\\
No&Data set& $\chi^2_{min}$/$N_F$& $\Omega_m$& $\Omega_{\Lambda}$&$\Omega_{\nu}$& $\Omega_b$  &$n_s$   & $h$  \\ [4pt]
\hline
&&&&&&&&\\
1&All observable data points are used                         &5.90/7&0.37&0.69&0.027&0.037&1.02&0.71\\
2&$\tilde P_{A+ACO}(k)$'s points are excluded                 &2.12/4&0.32&0.75&0.0  &0.039&1.00&0.70\\ 
3&$\tilde \ell_p$, $\tilde A_p$ are excluded                  &4.79/5&0.39&0.47&0.058&0.042&1.14&0.67\\
4&$\tilde V_{50}$ is excluded                                 &5.54/6&0.37&0.69&0.021&0.038&1.00&0.71\\ 
5&$\tilde\sigma_{8}\Omega_m^{\;\;\alpha_1}$ is excluded &4.58/6&0.45&0.61&0.052&0.039&1.03&0.69\\ 
6&$\tilde\sigma_{8}\Omega_m^{\;\;\alpha_2}$ is excluded &5.88/6&0.37&0.69&0.027&0.037&1.02&0.71\\ 
7&$\tilde\sigma_{8}\Omega_m^{\;\;\alpha_3}$ is excluded &4.72/6&0.38&0.68&0.028&0.038&1.01&0.70\\ 
8&All $\sigma_{8}$ tests are excluded                         &3.85/4&0.49&0.57&0.060&0.041&1.04&0.68\\ 
9&the first Ly-$\alpha$ test is excluded                          &5.46/6&0.42&0.65&0.048&0.039&1.02&0.70\\ 
10&The second Ly-$\alpha$ test is excluded                         &5.81/5&0.37&0.69&0.026&0.037&1.02&0.71\\ 
11&Both Ly-$\alpha$ tests are excluded                         &4.49/4&0.56&0.50&0.21 &0.042&1.04&0.67\\
12&The nucleosynthesis constraint is excluded                      &4.52/6&0.29&0.89&0.023&0.001&1.04&0.67\\ 
12& The direct constraint on $h$ is excluded                       &4.18/6&0.29&0.71&0.038&0.023&1.05&0.91\\ 
13&Both previous constraints are excluded                      &4.16/5&0.29&0.71&0.041&0.013&1.07&0.87\\ 
14&$\tilde V_{50}$, $\tilde\sigma_{8}\Omega_m^{\;\;\alpha_2}$ 
and $\Delta^2_{\rho}(k_p)$ are excluded                     &5.52/4&0.37&0.69&0.021&0.038&1.01&0.70\\ 
15&$\tilde\sigma_{8}\Omega_m^{\;\;\alpha_2}$ 
and $\Delta^2_{\rho}(k_p)$ are excluded                     &5.88/5&0.37&0.68&0.028&0.037&1.02&0.71\\ 
\hline
\end{tabular}
\end{center}
\end{table*}

 Comparing the results obtained without and with the SNIa constraint,
we conclude that the values of the fundamental cosmological parameters
$\Omega_m$, $\Omega_{\Lambda}$ and 
$\Omega_k$  determined by the observational characteristics of large
scale structure match the SNIa test very well. This can be interpreted
as independent support of the SNIa result in the framework of the
standard cosmological paradigm. However,
in order to elucidate how LSS data constraint cosmological parameters,
we analyze further only the model obtained without the SNIa constraint.
 
The best fit values of cosmological parameters determined by LSS
characteristics\footnote{We still include the direct measurement of
$h$ and the nucleosynthesis constraint in the analysis.} are  
$\Omega_m=0.37\pm0.06$, $\Omega_{\Lambda}=0.69\pm0.07$,
$\Omega_{\nu}=0.03\pm0.03$, $N_{\nu}=1$, $\Omega_b=0.037\pm0.009$,
$n_s=1.02\pm0.04$, and $h=0.71\pm0.09$. 
The CDM density parameter is $\Omega_{cdm} = 0.30\pm0.10$ and
 $\Omega_{k}=-0.06\pm0.13$. The neutrino content, which is compatible
with zero is very badly determined (100\% error). The obtained value
should be interpreted as an upper limit to the neutrino contribution. 
Below we will discuss this upper limit in more detail.

\begin{figure}
\epsfxsize=8truecm
\epsfbox{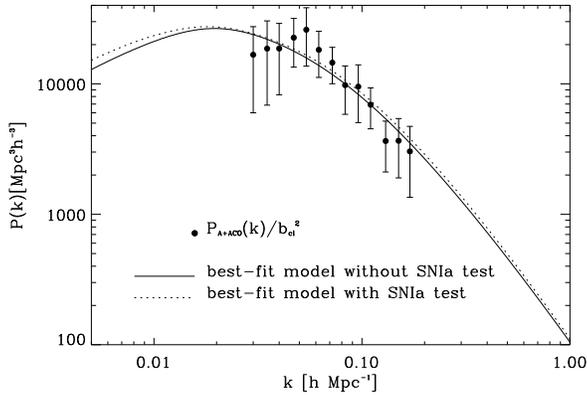}
\caption{The observed Abell-ACO power spectrum (filled circles) and
the theoretical spectra predicted by closed $\Lambda$MDM models with
parameters taken from Table~\ref{rez1} ($N_{\nu}=1$).}
\label{pkth}
\end{figure}

The value of the Hubble constant is close to the result by
Madore \etal~(1999) and Mould \etal (2000), somewhat higher than the
directly measured value 
given in Eq.~(\ref{Hubble}). The spectral index coincides with
the prediction of the simplest inflationary scenario, it is close
to unity.
The neutrino matter density $\Omega_{\nu}=0.03$
corresponds to a neutrino mass of $m_{\nu}=94\Omega_{\nu}h^2\approx1.4$ eV
but is compatible with 0 within $1\sigma$. The
estimated cluster bias parameter $b_{cl}=2.36\pm0.25$ fixes the amplitude
of the Abell-ACO power spectrum (Fig.~\ref{pkth}).  

Recently, it has been shown
\cite{nov99} that due to the large error bars, the position of the peak of
$\tilde P(k)$ at $k\approx 0.05$h/Mpc does not  influence
the determination of cosmological parameters significantly.
Only  the slope of the power spectrum on scales smaller than the
scale of the peak is relevant for cosmological parameters.
On the other hand, the relation of the peak  in $\tilde P_{A+ACO}(k)$
obtained from the space distribution of Abell - ACO clusters around us to
the matter density of the power spectrum of entire Universe is still under
discussion. Using numerical simulations, Retzlaff \etal~(1997)
 have shown that the pronounced peak in the spectrum (the fifth data
point in Fig.~\ref{pkth}) could be  purely due to cosmic variance. 
Therefore, it should not influence  cosmological parameters. 
In fact, the maximum of our fitting curve is at a different position, 
which shows that this peak position is not relevant for the present work.
The oscillation of the $\tilde P_{A+ACO}(k)$ around the 
best-fit $P(k)$  in Fig.~\ref{pkth} determined from all
observable data  on LSS reflects the real distribution of rich clusters of
galaxies in the vicinity of $\sim300$h$^{-1}$Mpc of our own galaxy only.
This is supported by similar features in spectra reconstructed from
the expanded
sample of Abell-ACO clusters \cite{mil00} and IRAS Point Source Catalog
Redshift Survey \cite{sau00,ham00}. 

\begin{figure}
\epsfxsize=8truecm
\epsfbox{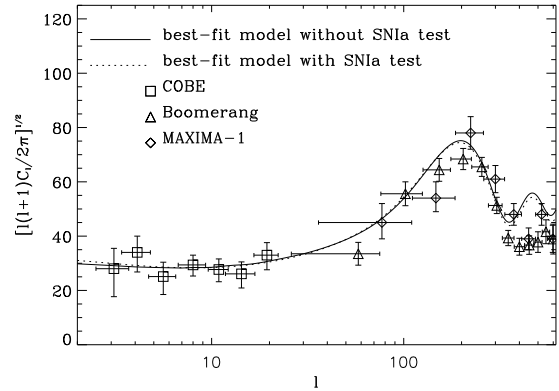}
\caption{The CMB power spectra predicted by best-fit $\Lambda$MDM models
with parameters  from Table~\ref{rez1} ($N_{\nu}=1$) and COBE DMR (Bennett et al. 1996), 
Boomerang (de Bernardis et al. 2000) and MAXIMA-1 (Hanany et al. 2000) experimental data.}
\label{cl}
\end{figure}

 Using CMBfast we have calculated the angular power spectra of CMB
temperature fluctuations for both best-fit models. Comparison with the 
COBE, Boomerang and MAXIMA-1 experiments are shown in Fig.~\ref{cl}.
 The CMB power spectrum predicted by both best-fit models
matches the data very well within the
range of the first acoustic peak. But it does not reproduce the
absence of a second peak inferred from the Boomerang and MAXIMA-1 data
at $\ell>350$.  This problem has been discussed intensively in literature
\cite{lan00,teg00b,hu00,emmmp00}. 
The lack of power in this range strongly favors models with more baryons than 
compatible with standard cosmological nucleosynthesis. The MAXIMA-1
data reduces the problem somewhat but does not remove it entirely
\cite{hu00}. However, as we shall discuss, the cosmological
parameters which match Boomerang and MAXIMA-1 data at high spherical
harmonics also strongly disagree with other
LSS constraints used here (see Subsection 4.8 below). Furthermore, 
the Boomerang, MAXIMA-1 and other CMB data
in this range do not match each other very well. This (and the amount
of work already published on this subject some of which is cited
above) prompted us to
ignore the problem of the second peak in the CMB anisotropy spectrum
in this work. Future flights of Boomerang and MAXIMA and/or the
future projects MAP and Planck will certainly decide on this very
important issue, but we consider it premature to draw very strong
conclusions at this point.

Finally let us mention some global characteristics of a Universe with our
best-fit cosmological parameters. Its age of $t_0=12.6$ Gyrs is in the
range of values determined from the age of globular clusters
\cite{cha98,car99}. The deceleration 
parameter is $q_0=-0.52$, in good agreement with the SNIa constraint
presented above leading to~\cite{per98} $\tilde q_0=-0.57\pm0.17$. The
original deceleration ($q>0$) changes into 
acceleration ($q<0$) at the redshift $z_d\approx0.55$. The 'equality epoch',  
$\rho_m(z_e)=\rho_{\Lambda}(z_e)$, corresponds to the redshift
$z_e\approx0.23$.

\subsection{The influence of different experimental data}

One important question is how sensitive our result responds to each
data point. 
To estimate this, we  exclude some data points from the search routine 
 and re-determine the best-fit parameters. The results of this
procedure are presented in Table~\ref{dds}.  In all cases when data on
the first acoustic peak are included $\Omega_m+\Omega_{\Lambda}\approx
1.06$, very slight positive curvature ($\Omega_k\approx-0.06$)
but compatible with flat,  i.e. the geometry is defined
mainly by the position of the first acoustic peak. 
The LSS data without CMB measurements prefer an open Universe with
$\Omega_k=0.14$ (4 row in the Table~\ref{dds}). The value of
$\Omega_m$ never exceeds 0.56, $\Omega_{\Lambda}$ is always larger
0.47 and in most cases $\Omega_{\Lambda}>\Omega_m$. The best-fit
values of the spectral index $n_s$ and $h$ for the different
observable data sets  are in the relatively narrow ranges of, 
0.99-1.14 and 0.67-0.72 respectively. The baryon content, $\Omega_b$
is fixed by the nucleosynthesis constraint. Without this constraint 
(12 row in Table ~\ref{dds}) $\Omega_b$ is lower, $\Omega_b
\approx0.001$, even below the value inferred from the luminous
matter in the Universe, $\Omega_{\rm lum}\sim 7\times10^{-3}$.

The hot dark matter content, $\Omega_{\nu}$, is reduced mainly by the 
Ly-$\alpha$ constraints but it is  poorly determined in all cases. 
If instead of or together with these Ly-$\alpha$ constraints we
use those by McDonald \etal~ (2000) which reduce the power at small
scales, then the best-fit value for the neutrino content is $\approx 0.07$.
But in this case the predictions for Ly-$\alpha$ constraints by Gnedin (1998) 
and Croft et al. (1998) are out of their $1\sigma$ ranges.
Moreover, the constraints by McDonald \etal~ (2000) are not in good
agreement  with other data, especially, Bahcall \& Fan (1998) and the
SNIa constraints. We have not included these constraints any further
in our determination of cosmological parameters.
Note however that the neutrino content is mainly constrained by the
Ly-$\alpha$ data. If both  Ly-$\alpha$ tests are excluded, the best
fit value of $\Om_\nu$ raises to 0.21! 

Excluding the direct measurement of the Hubble parameter from our
search procedure leads to a substantially larger value of $h\sim 0.91$
which is in disagreement with the direct determination. 

The comparison of the 1-st and 2-nd rows of Table~\ref{dds} shows that 
the Abell-ACO power spectrum prefers a slope of the matter power spectrum 
in the range $0.02\le k \le 0.1h$/Mpc $n\sim -1.5$ which results
in lowering $\Omega_{\Lambda}$ and introduces a small but non-zero
neutrino content.

The constraints  $\tilde\sigma_{8}\Omega_m^{\;\;\alpha_2}$ 
\cite{via99} and $\Delta^2_{\rho}(k_p)$ have
practically no influence on the determination of parameters 
(rows 6, 10 and 15) due to  their large error bars. They can be
removed from the  data set which reduces the number of effective degrees of
freedom  to $N_F=5$; this is important for the marginalization procedure.

\subsection{Marginalization}

 The next
important question is: which is the confidence limit for each parameter
marginalized over the others. The straight forward answer is the
integral of the likelihood function over the allowed range of all
the  other parameters.
But for a 7-dimensional parameter space this is computationally
time consuming.
Therefore, we estimate the 1$\sigma$ confidence limits for
all parameters in
the following way. By varying all parameter we determine
 the 6-dimensional $\chi^2$ hyper-surface which
contains  68.3\% of the total probability distribution.
We then project this hyper-surface onto  each axis
in parameter space. Its shadow on the parameter axes gives us the 1$\sigma$
confidence limits for the cosmological parameter under consideration.  The
1$\sigma$  confidence limits obtained  in this way for
$\Lambda$MDM models with one  sort of massive neutrinos are given in
Table~\ref{tabmax}. Including $\tilde\sigma_{8}\Omega_m^{\;\;\alpha_2}$  
and $\tilde\Delta^2_{\rho}(k_p)$ does not change the marginalized limits
significantly.  

\begin{table}
\caption{ The best fit values of all the parameters with
errors obtained by maximizing the (Gaussian) 68\% confidence contours
over all other parameters.\label{tabmax}}
\begin{center}
\begin{tabular}{|c|cc|}
\hline
 &\multicolumn{2}{c|}{central value and errors}\\
\cline{2-3}
parameter & without SNIa constraint&with SNIa constraint\\ [4pt]
\hline
&&\\
$\Omega_m$         &$0.37^{+0.25}_{-0.15}$    &$0.32^{+0.20}_{-0.11}$	 \\[4pt]
$\Omega_{\Lambda}$ &$0.69^{+0.15}_{-0.20}$    &$0.75^{+0.10}_{-0.19}$	 \\[4pt]
$\Omega_{\nu}$     &$0.03^{+0.07}_{-0.03}$    &$0.0^{+0.09}_{-0.0}$	 \\[4pt]
$\Omega_b$         &$0.037^{+0.033}_{-0.018}$ &$0.038^{+0.033}_{-0.019}$ \\[4pt]
$n_s$                &$1.02^{+0.09}_{-0.10}$    &$1.00^{+0.13}_{-0.10}$	 \\[4pt]
$h$                &$0.71^{+0.22}_{-0.19}$    &$0.70^{+0.28}_{-0.18}$	 \\[4pt]
$b_{cl}$           &$2.4^{+0.7}_{-0.6}$       &$2.2^{+0.8}_{-0.5}$	 \\[6pt]
\hline
\end{tabular}
\end{center}
\end{table}

\begin{figure}
\epsfxsize=8truecm
\epsfbox{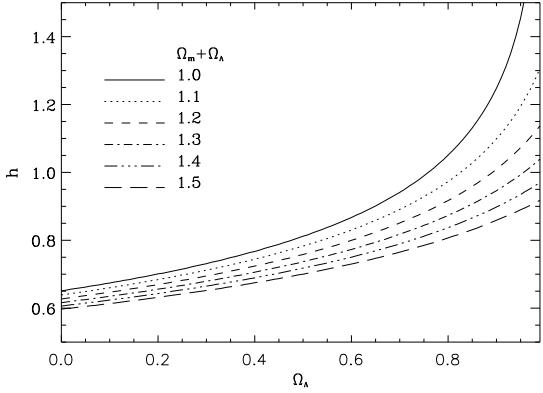}
\caption{The lines in the $\Omega_{\Lambda}-h$
plane corresponding to the lower limit on age of the Universe of 10Gyrs
established from oldest globular cluster for models with 
zero- and positive curvature. The range below the corresponding line
is allowed.}
\label{h_lambda}
\end{figure}

It must be noted that even though the upper $1\sigma$ edge for $h$ is
0.93 when marginalized over all other parameters for the data used here,
the resulting age of the Universe is still larger than the lowest value allowed
for  the age of the oldest globular clusters, $t_0\approx 10$ Gyrs if
$\Omega_{\Lambda}>0.72$. In the Fig.~\ref{h_lambda} we present the
constraints in the $\Omega_{\Lambda}-h$ plane given by the lower limit
for the age of the Universe, 10 Gyrs, for models with zero- and
positive curvature. The range above corresponding line is excluded by
this limit. Thus, the lower limit for the age of the Universe
additionally constrains  the confidence limits on the parameters, $h$
and $\Omega_{k}$ from above and on $\Omega_{\Lambda}$ from below. 

We have repeated the marginalization procedure including the SNIa test
(last column in Table~\ref{tabmax}). In
 this case we have to use all input data points (15 independent
measurements), since neglecting
$\tilde\sigma_{8}\Omega_m^{\;\;\alpha_2}$ and
$\Delta^2_{\rho}(k_p)$ does somewhat change the marginalized
limit. Hence, the number of degrees of freedom is $N_F=8$  (1$\sigma$
confidence limits corresponding to  $\chi^2\le15.3$).
The SNIa test reduces the confidence ranges of $\Omega_m$ and 
$\Omega_{\Lambda}$ in spite of the larger number of degrees of freedom, but it
results in somewhat wider $1\sigma$ error bars for the other
parameters due to the increase of $N_F$ and the low accuracy of the
added data points.

\subsection{The status of some subclasses of models}

The errors shown in Table~\ref{tabmax} define the range of each parameter 
within which by adjusting the remaining parameters a value of 
$\chi^2_{min}\le 11.8$ can be achieved. Of course, the values of the 
remaining parameters always lay within their corresponding
68\% likelihoods given in the Table~\ref{tabmax}. 
Models with vanishing $\Lambda$ are outside of this marginalized
$1 \sigma$ range of the best-fit model determined by the LSS
observational characteristics used here even without the SNIa
constraint (column 2). Let us investigate the status of these models
in more detail. For this, we set $\Omega_{\Lambda}=0$ as fixed parameter
and determine  the remaining parameters in the usual way. The minimal
value of  $\chi^2$ is $\chi^2\approx24$ with the following values for
the other parameters: $\Omega_m=1.15$, $\Omega_{\nu}=0.22$,
$N_{\nu}=3$, $\Omega_b=0.087$, $n_s=0.95$, $h=0.47$, $b_{cl}=3.7$
($\sigma_8=0.60$)). This model is outside the  2$\sigma$ confidence
contour of  the best-fit model for $N_{\nu}=3$ (Table ~\ref{rez1}
without SNIa test). The experimental data  which disagrees most with
$\Lambda=0$ is the data on the first acoustic peak. If we exclude it
from the experimental data set, $\chi^2_{min}\approx5.8$ for an open model
with following best-fit parameters: $\Omega_m=0.48$, $\Omega_{\nu}=0.12$,
$N_{\nu}=1$,  $\Omega_b=0.047$, $n_s=1.3$, $h=0.64$, $b_{cl}=2.5$ 
($\sigma_8=0.82$)). 
This model is inside the  1$\sigma$ confidence contour of the best-fit
$\Lambda$MDM model obtained without data on the first acoustic peak
(row 3 of Table~\ref{dds}). The reason for this
behavior is clear: the position of the 'kink' in the matter power
spectrum at large scales demands a 'shape parameter'
$\Gamma=\Omega_mh^2 \sim 0.25$ which can be achieved either by
choosing an open model or allowing for a cosmological constant. The
position of the acoustic peak which demands an approximately flat model then
closes the first possibility.

Results change only
slightly if instead of the Boomerang data we use Boomerang+MAXIMA-1 as
discussed in Section 4.1.
Hence, we  can conclude that the LSS observational 
characteristics together with the Boomerang (+MAXIMA-1) data on the
first acoustic peak already rule out 
zero-$\Lambda$ models at more than 95\% C.L.  and actually demand a
cosmological constant in the same bulk part as direct measurements. We
consider this a non-trivial consistency check!

Flat $\Lambda$ models in contrary, are inside the 1$\sigma$
contour of our best-fit model. Actually, the best fit flat model has 
$\chi^2_{min}\approx8.3$  and the best fit
parameters $\Omega_m=0.35\pm0.05$, $\Omega_{\Lambda}=0.65\mp0.05$, 
$\Omega_{\nu}=0.04\pm0.02$, $N_{\nu}=1$, $\Omega_b=0.029\pm0.005$, 
$n_s=1.04\pm0.06$, $h=0.81\pm0.06$, $b_{cl}=2.2\pm0.2$ 
($\sigma_8=0.96$) are close to our previous~\cite{nov00a} results
with a somewhat different observational data set. 

It is obvious, that flat zero-$\Lambda$ CDM and MDM models are ruled
out by the present experimental data set at even higher confidence
limit than by data without the Boomerang and MAXIMA-1
measurements in~\cite{nov00a}.

\subsection{Upper limits for the neutrino mass}

Since the neutrino content is compatible with zero, we determine an
upper limit for it. We first determine the marginalized
1$\sigma$, 2$\sigma$ and 3$\sigma$ upper limits for $\Omega_{\nu}$ for
different values of $N_{\nu}$. Using the best-fit value for $h$ at
given  $\Omega_{\nu}$, we can then determine  the corresponding upper
limit for the neutrino mass,  
$m_{\nu}=94\Omega_{\nu}h^2/N_{\nu}$. The results are presented in
Table~\ref{nu}.  For more species of massive neutrino the upper limit for
$\Omega_{\nu}$ is somewhat higher but $m_{\nu}$ is still lower for
each C.L. The  upper limit for 
$\Omega_{\nu}$ raises with the confidence level as expected. But the
upper limit for the mass grows only very little due to the reduction
of the best-fit value for $h$. The upper limit for 
the combination  $\Omega_{\nu}h^2/N_{\nu}^{0.64}$ is
approximately constant for all number species and confidence levels. 
The observational data
set used here establishes an upper limit for the massive neutrino
content of the universe which can be expressed in the form
$\Omega_{\nu}h^2/N_{\nu}^{0.64}\le0.042$  at 2$\sigma$ confidence 
level.The corresponding upper limit on the neutrino mass $m_{\nu}\le4$eV 
is close to the value obtained by \cite{cr99}.

\begin{table}
\caption{The upper limits for the neutrino content and mass (in eV) at
different confidence levels. 
\label{nu} } 
\begin{center}
\def\onerule{\noalign{\medskip\hrule\medskip}}
\medskip
\begin{tabular}{|c|cc|cc|cc|}
\hline
&\multicolumn{2}{c|}{1$\sigma$ C.L.}
&\multicolumn{2}{c|}{2$\sigma$ C.L.}
&\multicolumn{2}{c|}{3$\sigma$ C.L.}\\
$N_{\nu}$ &$\Omega_{\nu}$ &$m_{\nu}$ &$\Omega_{\nu}$ &$m_{\nu}$ &$\Omega_{\nu}$ &$m_{\nu}$   \\ [4pt]
\hline
1&0.10&3.65&0.13&3.96&0.18&4.04\\
2&0.15&2.79&0.21&3.06&0.29&3.35\\
3&0.20&2.40&0.27&2.67&0.35&2.78\\
\hline
\end{tabular}
\end{center}
\end{table}

\subsection{Limiting the tensor mode}

Up to this point we ignored uncertainties  in the COBE normalization.
 The statistical uncertainty of the fit to the
four-year COBE data, $\delta_h$, is 7\% (1$\sigma$) \cite{bun97} and
we want to study how this uncertainty
influences the accuracy of cosmological parameters which we determine.

Varying $\delta_h$ in the 1$\sigma$ range we found that the best-fit values of all 
parameters except $\Omega_{\nu}$  do not vary by more than 7\% from the values presented
in Table 1. Only $\Omega_\nu$, on which 1$\sigma$ errors are of the
order of 100\%, varies in a range of 20\% . These 
uncertainties are significantly  smaller than the standard errors
given in Table 1 and ignoring them is thus justified. (Including this
error raises our standard 1$\sigma$ errors from typically 10\% - 20\% to
11\% - 21\%.)

Our results depend on a possible tensor component only  via the COBE
data which enters our calculation through the normalization constant
$\delta_h$, in Eqns.~(\ref{pkz},\ref{anorm}). 
We can estimate the maximal contribution of a tensor mode in the
COBE $\Delta T/T$ data in the following way: we disregard the COBE 
normalization and consider $\delta_h$ as free
parameter to be determined  like the others.  Its best-fit value then
becomes $\delta_h^{LSS}=(2.95\pm2.55)\cdot 10^{-5}$ (for $N_{\nu}=1$),
while the best-fit values of the other parameters are 
$\Omega_m=0.40\pm0.08$, $\Omega_{\Lambda}=0.66\pm0.07$,
$\Omega_{\nu}=0.05\pm0.05$, $\Omega_b=0.038\pm0.010$, $n_s=1.14\pm0.31$, $h=0.71\pm0.09$ 
and $b_{cl}=2.4\pm0.3$. The best-fit value for density perturbation at horizon
scale from the 4-year COBE data for this set of parameters is
larger then the best-fit value determined from
LSS characteristics, $\delta_h^{COBE}=4.0\cdot10^{-5}>\delta_h^{LSS}$. This
means that COBE $\Delta T/T$ data may contain a non-negligible tensor
contribution. The most likely value of its fraction is given by 
$T/S=(\delta_h^{COBE}-\delta_h^{LSS})/\delta_h^{LSS}$. This value is
$T/S=0.36$ for the
corresponding best-fit values of $\delta_h^{COBE}$ and $\delta_h^{LSS}$
from  the Boomerang data alone  and $T/S=0.18$ 
from the combined Boomerang + MAXIMA-1 data. Since the standard error is
rather large, $\approx90\%$, we determine upper confidence limits for $T/S$ 
by marginalizing $\delta_h^{LSS}$ over all the other parameters like
we did for the neutrino content (see subsection 4.6). We then obtain
$T/S<1$ at 1$\sigma$ C.L. and $T/S<1.5$ at 2$\sigma$ C.L. from the
Boomerang data alone for the  amplitude and position of the first
acoustic peak. If we use the combined Boomerang + MAXIMA-1 data 
these limits are somewhat lower, 0.9 and 1.3 correspondingly, due to
the higher amplitude of the first acoustic peak measured by MAXIMA-1.
The 1$\sigma$ upper constraint on the tensor mode obtained recently by 
Kinney \etal~ (2000) from the Boomerang and MAXIMA-1 data on the CMB 
power spectrum for  the same class of models (T/S$<0.8$ in our definition)
is very close to the value obtained here.

\subsection{Comparison with other parameter estimations}

The cosmological parameters determined here from LSS+CMB data agree well
 with the values obtained by other methods (see e.g. the review by 
Primack (2000)). The marginalized $1\sigma$ ranges are still rather
large due to the large experimental errors, the large number of
parameters and the high  degree of freedom. But this does, of course, not
mean that an arbitrary set of parameters
within the marginalized ranges  matches the experimental data set
with an accuracy $\le 1\sigma$. 

We compare our best fit model with others found in the recent
literature by testing our data set as well as the Boomerang and
MAXIMA-1 data on the CMB power spectrum.  At first we calculate the
predictions of the following models for our data set  $(\Omega_m,
\Omega_{\Lambda},\Omega_b,n_s, h)={\bf P}=(0.49, 0.56, 0.054, 0.92,
0.65)$ obtained by Lange \etal~(2000) as best-fit model for the
Boomerang and LSS data (denoted there as model P9);
${\bf P}=(0.68, 0.23, 0.07, 1, 0.6)$ obtained by Balbi \etal~(2000) as
best-fit model to the MAXIMA-1 and COBE DMR data; ${\bf P}= (0.35,
0.65, 0.036, 0.95, 0.8)$  obtained by~Hu \etal~(2000) as best-fit
model to the Boomerang + MAXIMA-1 data on the first, second and third
acoustic peaks; ${\bf P}=(0.3, 0.7, 0.045, 0.975, 0.82)$ obtained by
Jaffe \etal~(2000) as best-fit model to the Boomerang + MAXIMA-1 +
COBE  data on the CMB power spectrum; and the "concordance" model by
Tegmark \etal~(2000) which favors
${\bf P}=(0.38, 0.62, 0.043, 0.91, 0.63)$. Some authors give several sets
of parameters obtained for different priors or by including different
data sets, we take the one from which we obtain a  minimal $\chi^2$
for our data set. All these models have no massive neutrino
component, no tensor mode and reionization is either not included or can be
neglected. The predictions of cosmologies with the above parameters
for the data  considered in this work are presented in Table~\ref{comp}. 
\begin{table*}
\caption{Theoretical  predictions for the observational values by
best-fit models from the literature:
 A (Lange et al. 2000), B (Balbi et al. 2000), C (Hu et al. 2000), 
D (Jaffe et al. 2000), E (Tegmark et al. 2000).\label{comp} }
\begin{center}
\def\onerule{\noalign{\medskip\hrule\medskip}}
\medskip
\begin{tabular}{|ccccccc|}
\hline
&&&&&\\
&&\multicolumn{5}{c}{Predictions}\\
\cline{3-7}
Characteristics& Observations&A &B &C &D &E \\ [4pt]
\hline
&&&&&&\\
$\ell_p$                       &197$\pm$6        &206   &231 &206   &213    &225    \\
$A_p$                          &69$\pm$8         &57    &68  &63    &72     &62     \\
$V_{50}$, km/s                 &375$\pm$85       &280   &310 &303   &293    &239    \\
$\sigma_8\Omega_m^{\alpha_1}$  &0.60$\pm$0.022   &0.64  &0.91&0.68  &0.58   &0.43    \\
$\sigma_8\Omega_m^{\alpha_2}$  &0.56$\pm$0.095   &0.62  &0.93&0.65  &0.65   &0.42   \\
$\sigma_8\Omega_m^{\alpha_3}$  &0.8$\pm$0.1      &0.70  &0.96&0.79  &0.73   &0.49    \\
$\sigma_F$                     &$2.0\pm .3$      &1.7   &2.8 &2.2   &1.9    &1.1     \\
$\Delta^2_{\rho}(k_p)$         &$0.57\pm 0.26$   &0.51  &1.21&0.81  &0.62   &0.25   \\
$n_{p}(k_p)$                    &$-2.25\pm0.2$   &-2.25&-2.15&-2.21 &-2.22  &-2.30   \\
$h$                            &0.65$\pm$0.10    &0.65  &0.60&0.80  &0.82   &0.63   \\
$\Omega_bh^2$                  &0.019$\pm$0.0012 &0.023 &0.025&0.023&0.030  &0.02    \\
$\Omega_m-0.75\Omega_{\Lambda}$&-0.25$\pm$0.125  &0.07  &0.51 &-0.13&-0.23 &-0.09   \\
\hline
$\chi^2$                       &                 &27    &285  &39   &105    &106    \\
\hline
\end{tabular}
\end{center}
\end{table*}
\begin{table*}
\caption{The $\chi^2$ deviation of theoretical  predictions
for the CMB power spectrum from experimental results for the models 
in Table~\ref{comp} and for our best-fit model. The first number represents
the the value of $\chi^2$  for the CMB power spectrum in the range of
the first acoustic peak, $50\le \ell \le 375$, the second number is
for the entire range $50\le \ell \le 750$. Clearly our model
parameters are in serious disagreement with the experimental CMB data
beyond the first acoustic peak. \label{chi}}
\begin{center}
\def\onerule{\noalign{\medskip\hrule\medskip}}
\medskip
\begin{tabular}{|ccccccc|}
\hline
&&&&&\\
&\multicolumn{6}{c}{$\chi^2$}\\
\cline{2-7}
Experiment&A &B &C &D &E & our best-fit model\\ [4pt]
\hline
&&&&&&\\
Boomerang                     &7.3/12.6   &77.2/96.7   &6.1/12.8     & 18.3/24.5     & 13.6/24.5    & 11.3/108.5  \\
MAXIMA-1                      &16.9/18.7  &4.6/11.4    &15.2/17.0    & 10.3/11.7     & 16.2/21.6    & 11.1/48.5   \\
\hline
\end{tabular}
\end{center}
\end{table*}

The $\chi^2$ presented in last row includes also 
$\chi^2_{A+ACO}=\sum_{i=1}^{13}\left({P_{A+ACO}(k_i)-b^2_{cl}P(k_i)\over \Delta P_{A+ACO}(k_i)}\right)^2$
which is small due to the cluster bias, $b_{cl}$, which is considered as
free parameters in each model. 
In spite of the fact that all parameters of each model are within the
marginalized 1$\sigma$  ranges of the parameters of our best-fit
model, the total value of $\chi^2$ for the entire parameter sets rules
out all the  models at more than $2\sigma$ confidence level. 
Table~\ref{comp} indicates the crucial tests. Models A and C are ruled
out mainly by the nucleosynthesis constraint and the first $\sigma_8$
test (cluster mass function). Model B strongly disagrees with all $\sigma_8$
tests ($14\sigma$, $2.6\sigma$ and $1.6\sigma$ correspondingly), both
Ly-$\alpha$ tests ($2.6\sigma$ and $2.5\sigma$), the nucleosynthesis
constraint ($5.2\sigma$) and the data on the location of the first
acoustic peak ($5.6\sigma$). Moreover, models A and B do not match
the SNIa test which we have not included into $\chi^2$.
Model D strongly disagree with nucleosynthesis constraint (9.4$\sigma$)
and the Boomerang data on the 
location of the first acoustic peak ($2.6\sigma$). Model E does not
match first and third  $\sigma_8$ tests (at $5.1\sigma$ and
$2.4\sigma$ respectively), the first Ly-$\alpha$ test 
(at $2.4\sigma$) and the data on the location of the first acoustic peak
($3.2\sigma$).  The latter is due to the fact that the MAXIMA-1 peak
position is more than $1\sigma$ away from the peak position derived by
the Boomerang data alone.

We now  calculate the CMB power spectra for these models 
using CMBfast (version 3.2) and compare them with 
the experimental data from Boomerang~\cite{ber00} and
MAXIMA-1~\cite{han00}. The $\chi^2$ deviations for all models
including our best-fit model are presented in the Table~\ref{chi}. The
first number indicates the $\chi^2$ for the range of the first acoustic peak,
$50\le \ell \le 375$ (7 and 5 data points for the Boomerang and
MAXIMA-1 experiments respectively), for the second number we have used
the entire range, $50\le \ell \le 750$  (12 and 10 data points for the
Boomerang and MAXIMA-1 experiments respectively). 
In the range of the first acoustic peak our model fits as well as the
other models, but the observed power spectrum at higher spherical harmonics 
is not reproduced by our model as we mentioned above. 

Therefore, models which match the Boomerang and/or MAXIMA-1 CMB
power spectrum at high spherical harmonics (in the range of second and
third acoustic peak) disagree with some of the 
$\sigma_8$, Ly-$\alpha$ and/or the nucleosynthesis constraints. And
vice versa, model which match very well the LSS observational characteristics 
predicts the CMB power spectrum which disagrees with measurements by 
Boomerang and MAXIMA-1 on very small scales. 
The resolution of this problem can go in several
directions. If the Boomerang and MAXIMA-1 measurements are confirmed,
nucleosynthesis may have been more complicated than assumed for the
constraint used in this work~\cite{emmmp00}. An other problem may be
the cluster mass function constraint which is exponentially sensitive
to the value of $\sigma_8$ and might be too constraining, especially
in view of all the uncertainties in the theory of cluster
formation. Therefore, our constraint 
$\sigma_8\Omega_m^{\alpha_1}=0.60\pm0.022$ has to be taken with a
grain of salt and its incompatibility with, e.g. the CMB data may also
hint to a problem in the theory of cluster formation. Last but not
least, if inconsistencies in the determination of cosmological
parameters persist even after a serious improvement of data, e.g. with
the Sloan digital sky survey, this may hint that the correct model is
not within the class considered. If we want to fit a snail within the
class of all known mammals by $\chi^2$ minimization (or by a much more
sophisticated method), we never obtain a very convincing fit.

\section{Conclusions}

The main observational characteristics on  LSS together with recent
data on the amplitude and location 
of the first acoustic peak in the CMB power spectrum, and the
amplitude of the primordial power spectrum inferred by the COBE DMR
four year data prefer a $\Lambda$MDM model with the following parameters: 
$\Omega_m=0.37^{+0.25}_{-0.15}$,    
$\Omega_{\Lambda}=0.69^{+0.15}_{-0.20}$,    
$\Omega_{\nu}=0.03^{+0.07}_{-0.03}$,    $N_{\nu}=1$,
$\Omega_b=0.037^{+0.033}_{-0.018}$, 
$n_s=1.02^{+0.09}_{-0.10}$,    
$h=0.71^{+0.22}_{-0.19}$,    
$b_{cl}=2.4^{+0.7}_{-0.7}$
(1$\sigma$ marginalized ranges).       
 
The central values correspond to a slightly closed ($\Omega_k=-0.06$)
$\Lambda$MDM model with one sort of 1.4eV neutrinos. These neutrinos
make up about 8\% of the clustered matter, baryons are 10\%
and the rest (82\%) is in a cold dark matter component. The energy
density of clustered matter corresponds to only 35\% of the total
energy density of matter plus vacuum which amounts to $\Omega= 1.06$. 
The massive neutrino content is compatible with zero and we have established
an upper limit in the form of $\Omega_{\nu}h^2/N_{\nu}^{0.64}\le0.042$
at 2$\sigma$ confidence level. The upper 2$\sigma$ limit
for the neutrino  mass is 4.0eV.

If COBE normalization is disregarded, the best-fit value of the density
perturbation at horizon scale is $\delta_h^{LSS}=(2.95\pm2.55)\cdot
10^{-5}$ while the best-fit values of the other parameters are 
$\Omega_m=0.40$, $\Omega_{\Lambda}=0.66$, $\Omega_{\nu}=0.05$, $N_{\nu}=1$,
$\Omega_b=0.038$, $n_s=1.14$, $h=0.71$ and $b_{cl}=2.4$. Comparison it with 
the best-fit value to the COBE 4-year data $\delta_h^{COBE}$ gives an estimate
for the contribution of a tensor mode to the COBE DMR data:
$T/S=0.36^{+0.64}_{-0.36}$ from the Boomerang data on the first acoustic peak 
 and $T/S=0.18^{+0.72}_{-0.18}$ ($1\sigma$ confidence limits) when the
combined Boomerang+MAXIMA-1 data are used. The upper limits on $T/S$ at 
$2\sigma$ C.L. for these two cases are 1.5 and 1.3 respectively.

The values for  the matter density $\Omega_m$ and the
cosmological constant $\Omega_{\Lambda}$ for the best-fit model are
close to those  deduced from the SNIa test. Including this test in the
observational data set, results to a somewhat larger value of
$\Omega_{\Lambda}$  (7\%) and  slightly lowers $\Omega_m$. 

The observational characteristics of large scale structure together
with the  Boomerang (+MAXIMA-1) data on the first acoustic peak rule out 
zero-$\Lambda$ models at more than  $2\sigma$ confidence limit.

\section*{Acknowledgments} 
It is a pleasure to acknowledge stimulating discussions with Stepan
Apunevych, Pedro Ferreira, Vladimir Lukash, Max Tegmark and Matts Roos.
This work is part of a project supported by the
Swiss National Science Foundation (grant NSF 7IP050163).
B.N. is also grateful to Geneva University for hospitality.

\end{document}